\documentclass[prl,showpacs,twocolumn]{revtex4}
\usepackage{graphicx}
\newcommand{\mng}{Mn$_5$Ge$_3$ $\:$}
\begin{document}

\title{First-principles characterization of ferromagnetic
Mn$_5$Ge$_3$ for spintronic applications}
\author{S. Picozzi,   A. Continenza}
\affiliation{Center for Scientific and Technological Assistance
to Industries (CASTI) - Istituto Nazionale di Fisica della Materia (INFM) and
Dipartimento di Fisica,
Universit\`a degli Studi di L'Aquila, I--67010 Coppito (L'Aquila),
Italy }
\author{A.J.Freeman}
\affiliation{Department of Physics and Astronomy, Northwestern University,
 Evanston, Il, 60208 USA}
 
\begin{abstract}
In the active
search for potentially promising   
candidates for spintronic applications, we focus
on the intermetallic ferromagnetic \mng compound and perform
accurate first-principles FLAPW calculations within  density functional theory. 
Through a careful investigation of the bulk electronic and magnetic structure, 
our results for the total magnetization, atomic magnetic moments, metallic conducting
character
and hyperfine fields are found to be in good agreement with experiments, and are elucidated
in terms of a hybridization mechanism and exchange interaction. 
In order to assess the potential of this compound for spin--injection
purposes, we calculate  Fermi velocities and degree of
spin--polarization; our results predict a rather high
spin--injection efficiency in the diffusive regime along the hexagonal
$c$-axis.
Magneto-optical properties, such as
$L_{2,3}$ X-ray magnetic circular dichroism, are also
reported and await comparison with experimental data. 
\end{abstract}

\maketitle

\section{Introduction}
Mn--doped Ge has recently been proposed as a promising candidate in the
challenging field of diluted magnetic semiconductors 
(DMS)\cite{science,rmpdassarma}, 
which aims at combining information 
logic and storage. For example,
epitaxial single crystal films of Mn$_x$Ge$_{1-x}$ ($x<$ 8-10$\%$)
grown on GaAs(001) and Ge 
were found to exhibit Curie temperatures over
the range 25 to 116 K, combined with a $p$-type semiconducting
behaviour\cite{park}.  Many efforts are presently devoted toward
increasing the transition temperature up to or above room temperature. Within
this framework, one possible way is to increase the 
concentration of  magnetic impurities. However,
one of the key issues in 
DMS is indeed the solubility of Mn in the semiconducting host: it is well known
that beyond a certain critical
Mn concentration (typically of the order of a few percent
in III-V hosts), a tendency toward clustering and
phase separation occurs, thereby limiting the homogeneity and
 growth control that are strictly required for materials to be used in
 spintronic applications. 
This tendency was observed also during Mn-alloying of Ge samples:  Mn$_x$Ge$_y$
precipitates were detected during out-of-equilibrium growth\cite{parkapl}.

Intermetallic compounds of Mn and Ge occur in several different
stoichiometries and crystallographic phases\cite{matsui}, most of which are
antiferromagnetic or ferrimagnetic with rather low ordering temperatures.
However, \mng shows ferromagnetism with a Curie temperature of $\sim$300 K, 
along with a uniaxial magnetic anisotropy along the $c$ axis of the
hexagonal crystal structure (see below)\cite{tawara,brown,aplerwin}. 
Ferromagnetic \mng
thin
films grown epitaxially on Ge(111) by means of
 solid--phase epitaxy\cite{aplerwin} 
exhibited metallic conductivity and strong ferromagnetism up to 296 K --- thus holding
out promise for use in spin injection. Moreover, very recently  
point contact Andreev
reflection spectroscopy was used to measure the spin--polarization of \mng
epilayers\cite{pcarerwin} and the results were compared
with calculated values within the density functional theory.
 The discrepancy between the experimental and predicted
spin--polarization  was attributed to the extreme sensitivity of
calculated results to the crystallographic structure, as well as to possible
Mn deficiencies in \mng samples.\cite{pcarerwin}
Finally, it was shown experimentally,\cite{gajdzik}
  upon C
doping (with carbon
 interstitially incorporated into the voids of Mn octahedra of the
\mng compound), that the Curie temperature, $T_C$, 
dramatically increased: Mn$_5$Ge$_3$C$_x$
films for C concentration $x \geq$ 0.5 showed $T_C\sim$ 680 K.

So far, very little is known theoretically
about \mng; in particular, a careful
investigation from first-principles of the magnetic interactions and
chemical bonding between Mn and metalloid atoms is still lacking.
In this work, we perform a comprehensive study of \mng within
density functional theory; in
particular, in Sec.\ref{tech} we report the technicalities related to the
structure and to the computational approach. The electronic structure, as well as the
related magnetism, is discussed in
Sec.\ref{elec}, in terms of band structure, orbital and spin
magnetic moments, hyperfine fields and magnetic-circular dichroism spectra. 
Conclusions are drawn in Sec.\ref{concl}.

\section{Structural and computational details}
\label{tech}

Our calculations were performed using one of the most accurate available
density functional theory (DFT) methods, namely the all--electron
full-potential linearized augmented plane wave (FLAPW)
\cite{FLAPW}  approach. The generalized gradient approximation (GGA) according
to the Perdew-Becke-Erzenhof scheme\cite{pbe}
 was used for the exchange-correlation ($XC$)
potential. This choice was suggested by the more accurate treatment of this
exchange--correlation functional for magnetic compounds with respect to the
local spin-density approximation (LSDA)\cite{pnic}; however, in order to
test the resulting effects of a different
 ($XC$) parametrization and for the evaluation of the hyperfine fields, 
 we also performed some calculations
using the von Barth-Hedin\cite{vbh} functional within 
the local spin density approximation (LSDA).
We used  plane waves with wave vector up to $K_{max}$ = 3.8 a.u.,
leading to about 1500 basis functions, whereas
 for the potential and the charge density
we used an angular momentum expansion with $l_{max}$ $\leq$ = 8.  
The Brillouin zone
sampling was performed using 60 special $k$--points in the irreducible
wedge, according to  the
Monkhorst-Pack scheme \cite{MP}.
 The muffin tin radii, $R_{MT}$, for Mn and Ge were chosen equal
to 2.37 a.u. and 2.0 a.u., respectively. In order to evaluate the effects
 of the orbital contribution to the magnetic moments, the calculations 
were performed
with and
without the spin--orbit coupling (SOC) included in the Hamiltonian\cite{macsoc}. 

For the purpose of calculating the electronic group velocity {\bf v}({\bf k})=
(1/$\hbar$)[$\partial \varepsilon$({\bf k})/$\partial${\bf k}], the eigenenergies
$\varepsilon$({\bf k}) over a set of 150 {\bf k} points were used for a spline
fitting of the bands over the Brillouin zone \cite{dale}. The resulting
interpolating Fourier series was then used to calculate the
required energy derivative.
A similar approach has been followed to calculate the electronic plasma frequency:
\begin{equation}
\omega_{p\alpha\beta}^2 = \frac{4\pi e^2}{\Omega} N(E_F) 
<v_{\alpha}({\bf k})v_{\beta}({\bf k})>
\end{equation}
where $<>$ denotes the Fermi surface average. Given the hexagonal symmetry,
 the quantities, 
 $\omega_{pxx}$ = $\omega_{p\parallel}$
and $\omega_{pzz}$ = $\omega_{p\perp}$,
will be evaluated.

According to Forsyth and Brown\cite{brown},  intermetallic \mng   has an
hexagonal
 crystal
structure of D8$_8$ type (space group P6$_3$/mcm), with experimental
cell dimensions at
room temperature $a$ =
7.184 \AA $\:$ and $c$ = 5.053 \AA. The atomic positions are:

\begin{center}
Mn1 in 4(d) site: $\pm(\frac{1}{3},\frac{2}{3},0; 
\frac{2}{3},\frac{1}{3},\frac{1}{2})$

Mn2 in 6(g) site: $\pm(x,0,\frac{1}{4}; 0,x,\frac{1}{4}; 
-x,-x,\frac{1}{4})$ with $x$
= 0.2397

Ge in 6(g) site: $\pm(x,0,\frac{1}{4}; 0,x,\frac{1}{4}; 
-x,-x,\frac{1}{4})$  with $x$
= 0.6030

\end{center}

\begin{figure} 
\includegraphics[scale=0.6]{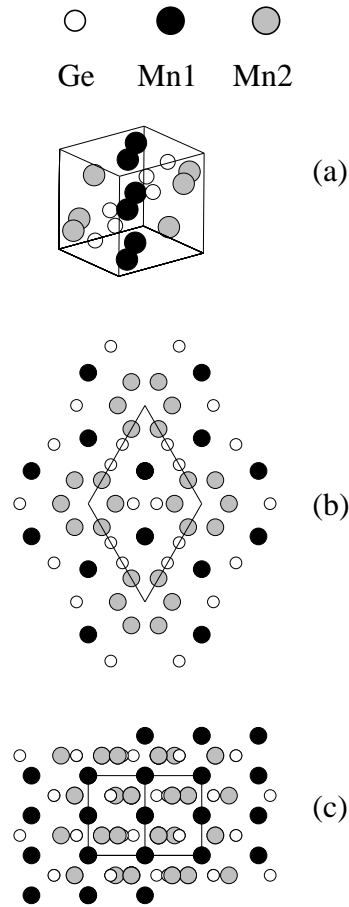}
\caption{(a) Perspective, (b) top and (c) side views of Mn$_5$Ge$_3$. Black,
white and
grey spheres denote Mn1, Ge and Mn2 atoms. The unit cell is also shown.}
\label{struc}
\end{figure}

Starting with the experimental
 equilibrium parameters, we checked that
the calculated 
internal atomic forces were negligibly small and that the minimum total
energy was obtained for the $a$ and $c$ value reported in 
Ref.\onlinecite{brown}.
This confirmed that the FLAPW method as well as  the GGA parametrization
   accurately
reproduce the experimental structural properties for compounds with a high
concentration of magnetic
atoms. 
In Figure \ref{struc}, we show the perspective, top and side views of the crystal.
It is evident that there are two different atomic
planes perpendicular to the [001] direction: the first  contains only
Mn1 atoms (at $z$ = 0 and $z$ = $c$/2, equivalent by symmetry) 
forming an hexagonal two-dimensional
lattice; the second  contains Mn2 and Ge
atoms (at $z$ = $c$/4 and $z$ = 3$c$/4, equivalent by symmetry).
We recall that in the complex \mng  structure, Mn1 and Mn2
atoms have different coordinations; 
in particular\cite{brown}, the nearest neighbours ($nn$)
of each Mn atoms are arranged as:
\begin{itemize}
\item Mn1 has ({\em i}) 
two (six) Mn1 (Mn2) nearest-neighbors at 2.522
(3.059) \AA $\:$ and   ({\em ii}) six Ge at 2.534 \AA; 
\item Mn2 has two Mn2, four Mn2 and four Mn1
at 2.976, 3.051 and 3.059 \AA, respectively and  ({\em ii}) two Ge, one Ge and two Ge
at 2.482, 2.606 and 2.762 \AA, respectively.
\end{itemize}

In this configuration, the enthalpy of formation,  $\Delta\:H_f$,
of  \mng  is evaluated with respect to the stable
phases of Mn ([001]-ordered antiferromagnetically fcc) and Ge (in the zincblende
phase). Our GGA calculated value, $\Delta\:H_f$ = 0.84
eV/formula--unit ({\em i.e.} $\sim$0.1 eV/atom), shows that  \mng
is a quite stable compound. 


\section{Electronic and magnetic properties}
\label{elec}
\subsection{Density of states and band structure}

\begin{figure} 
\includegraphics[scale=0.6]{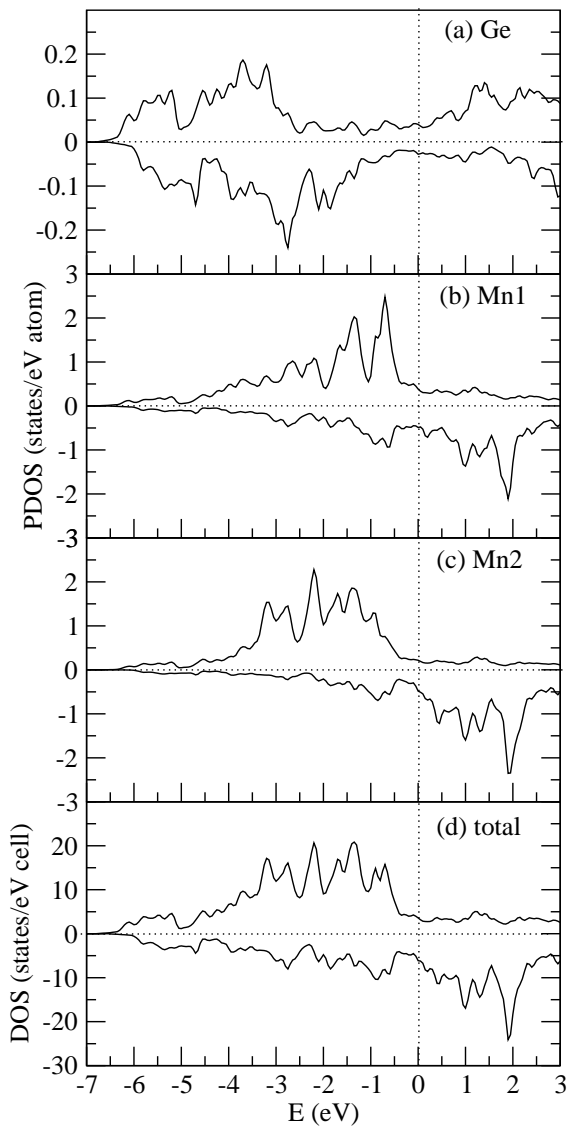}
\caption{PDOS of (a) Ge, (b) Mn1 and (c) Mn2 the total density of states is
shown in panel (d). Majority (minority) spin components are shown in the
positive (negative) $y$-axis. The $E_F$ is set to zero of the energy
scale.}
\label{pdos}
\end{figure}
 The projected density of states   (PDOS)
of the three different
atomic types forming the unit cell
and the total DOS are reported in Figure \ref{pdos}. As for its conducting character,
 \mng shows  strongly 
metallic behavior in both the minority and majority spin
components: this is consistent with recently reported electrical resistivity
experiments\cite{aplerwin}. 
The Ge atom
shows slightly different PDOS for up and down spins,
consistent with the small negative magnetic moment (see below).  
In the energy range considered, the largest
 contribution is due to $p$ states.
On the Mn sites, as expected, contributions from $s$ and
$p$ states (not shown) are negligible
and the PDOS is  essentially dominated by the 3$d$ states. 
As a difference with the case of Mn impurities in Ge\cite{stroppa}, 
we here have a
much smaller
hybridization between Ge $p$ and Mn $d$ states. In fact, we find the Ge $p$ states at higher
binding energy while the Mn $d$ dominate the region close to
the Fermi level, showing a quite large Mn--Mn interaction.

Differences in the
PDOS of Mn1 and Mn2 are particularly evident in the majority spin occupied states;
in particular, from the analysis of the peaks on different atoms located
at the same energies, we can infer that: {\em i})
the Mn1 features at $\sim$-0.7 eV and in--between -2 and -3 eV
are due to Mn1--Mn1 interactions; {\em ii}) the feature at  
$\sim$-1-2 eV is  due to Mn1-Mn2 interactions;  {\em iii})
the Mn2 feature at
-3.5-2.5 eV can be ascribed to Mn2-Mn2 interactions. The high--binding energy
range ($<$-3.5 eV) shows hybridization of both Mn atoms with Ge. Similarly, 
minority states for binding energies greater than 1.3 eV show common features
for Mn1, Mn2 and Ge atoms, whereas the feature at around -1-0.5 eV results from
Mn1-Mn2 hybridization. The unoccupied states, of interest for the discussion of
  magneto-optical properties (see below), are largely due to the minority spin
component and only show minor differences between Mn1 and Mn2.

\begin{figure}
\includegraphics[scale=0.6]{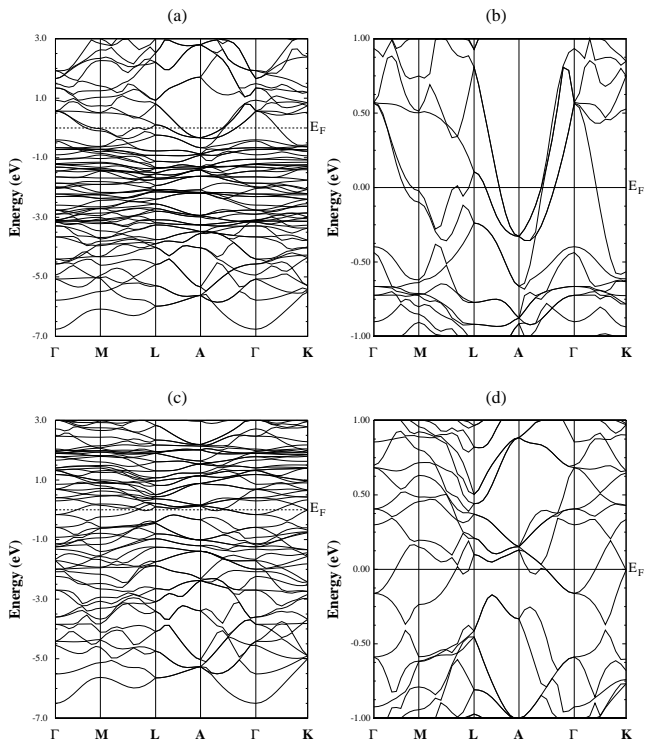}
\caption{(a) Majority and (c) minority band structure along the
main symmetry lines. Panels (b) and (d) show a blow-up around the Fermi
level. The $E_F$ is set to zero of the energy
scale.}
\label{bande}
\end{figure}

 The band structure for the majority and minority
spins is shown in Figure \ref{bande}. We remark that the levels around  $E_F$ 
(see Figure \ref{bande} (b) and (d)) are rather
dispersed for the majority spin channel, whereas they are more localized in the
minority spin component; moreover, a non--dispersed region is evident for higher
binding energies in the majority spin
band structure (in the energy range from -0.7
eV  to -3.5 eV), as well as in the unoccupied minority spin band structure. 
The dispersion around $E_F$ in the up-spin component
confirms the strong hybridization between Mn1
and Mn2 states, and between Mn $d$ and $p$ Ge states. 
Roughly speaking, in fact, as already pointed out for the DOS,  
the region at higher binding energies ({\em i.e.} in the
energy range between -7 eV and -1.5 eV) is basically due to a large
contribution from Ge; on the other hand, the levels around $E_F$ are
basically due to both Mn1 and Mn2 in the minority spin channel, whereas a $p$ Ge
contribution hybridized with Mn states is evident in the majority spin bands.

\subsection{Magnetic moments, spin and charge density}

\begin{table}
\caption{Magnetic moments within the atomic spheres and total magnetization
per unit cell (in Bohr magnetons). The first two lines show the spin magnetic
moments ($\mu^s_{NOSOC}$) as obtained from  calculations
without SOC within LSDA and GGA; the  third, fourth and fifth lines show the 
spin ($\mu^s_{SOC}$), orbital  ($\mu^l_{SOC}$) and
total ($\mu^t_{SOC}$) magnetic moments as obtained 
including the SOC self-consistently  within GGA. 
The magnetic moments of the atomic
species are calculated within their muffin-tin spheres, whereas the total
contribution includes the contribution from the interstitial region.
Experimental values ($\mu^t_{exp}$) from 
Refs.\protect\onlinecite{aplerwin,brown}
are shown in the last row.}
\begin{tabular}{|c|c|c|c|c|}\hline \hline
 & Ge & Mn1 & Mn2 & Total \\ \hline \hline
$\mu^s_{NOSOC}$ (LSDA) & -0.09 & 2.14 & 3.11 & 26.7 \\ \hline \hline
$\mu^s_{NOSOC}$ (GGA)  & -0.11 & 2.28 & 3.22 & 27.5 \\ \hline \hline
$\mu^s_{SOC}$ (GGA) & -0.11 &  2.07&  3.12& - \\ \hline
$\mu^l_{SOC}$ (GGA) & $~$0 & 0.05 &  0.035 & - \\ \hline
$\mu^t_{SOC}$ (GGA) & -0.11 & 2.12 & 3.16 &  25.9\\ \hline \hline
$\mu^t_{exp}$  & - & 1.96$^a$ & 3.23$^a$ & 26$^b$ \\ \hline \hline
\end{tabular}

\vspace{0.5cm}

$a$. Ref.\onlinecite{brown}

$b$. Ref.\onlinecite{aplerwin}
\label{tabmom}
\end{table}

 The calculated
total magnetization and the magnetic
moments of the different atomic species are compared with their corresponding
experimental values and reported in Table \ref{tabmom}. 
We show the magnetic moments (both spin and orbital contributions)  
with and without the
inclusion of SOC. 
As experimentally reported\cite{brown}, 
all the moments are ferromagnetically oriented along the
 crystallographic $c$ axis. A comparison between  LSDA and
GGA does not show any significant differences as far as the general
magnetization distribution over different atomic sites
is concerned; however, LSDA predicts lower
atomic magnetic moments (by about 0.1-0.15
$\mu_B$) than GGA, resulting  
in a LSDA total magnetic moment per unit cell that differs by almost
one Bohr magneton
with respect to its GGA counterpart.

According to previous neutron scattering experiments\cite{forsythold}, 
the magnetic structure of  \mng is found to reveal two Mn
sublattices (Mn1 in a four-fold and Mn2 in a six-fold position)
with different magnetic moments. It was suggested that Mn2 carries the
larger moment, in agreement with zero-field NMR measurements\cite{jackson}.
Our results are in excellent
quantitative and qualitative agreement with these results. 
As suggested in Ref.\onlinecite{brown}, the lower Mn1 magnetic moment is 
 due to
the different Mn coordination and to direct
Mn--Mn interactions at a rather short distance (recall, in fact, that every 
Mn1 atom in \mng has 2 Mn1 at a  distance of 2.52 \AA): 
 for a large
set of Mn intermetallic compounds, an analysis of the Mn magnetic 
moment vs the nearest neighbor distance\cite{brown}
showed that  below a ``threshold" separation of
3.1 \AA, a moment reduction of $\sim$ 2 $\mu_B$/\AA$\:$ 
per Mn neighbor occurred with respect to the
atomic value (5 $\mu_B$) of the Mn$^{2+}$ ion, therefore accounting for
 the small
magnetic moment at the Mn1 site.
 
 It is remarkable that the inclusion of the orbital
moments (which are not completely
negligible on both the Mn1 and Mn2 sites)
largely improves the agreement with experiment: the total
magnetic moment is in excellent agreement with the experimental saturation
magnetization as obtained by Kappel\cite{kappel}. Moreover, it was suggested by Forsyth
and Brown\cite{brown} that some {\em spatially diffuse} reverse magnetization 
(in contrast to
{\em localized} moments centered on the Mn atoms) exists in Mn$_5$Ge$_3$. 
This is
consistent with our calculated negative moment on Ge sites. As expected, the
Mn magnetic moments are due to $d$ states, whereas the small
induced moment on Ge is due to $p$ state polarization. 
In order to further
investigate this issue and to better show the bonding, 
we plot the charge  and spin density on two different planes perpendicular
to the [0001] axis. 
The 
charge density in Figure \ref{chfig} (a) shows fairly ``isolated" Mn1 atoms; a
somewhat stronger interaction occurs between Ge and Mn2 atoms, as shown by the
presence of charge in the bonding regions in Figure \ref{chfig} (c).

As regards deviations from spherical symmetry  in the spin density,  
we remark that Mn2 shows an almost 
spherical
shape, whereas the spin density around Mn1 is extended towards the six
surrounding Ge ligands. 
It is evident from Figure \ref{chfig}(d) that a negative spin-density
surrounds the metalloid Ge atoms, consistent with the negative magnetic
moment reported in Table \ref{tabmom}. Moreover, 
the {\em   diffuse} negative spin density shown in
 Figure \ref{chfig} (b)
and   (d)
 confirms the experimental results.\cite{brown}
\begin{figure}
\caption{(a) Valence charge density and (b) spin density in a plane
perpendicular to the [0001] direction and containing  Mn1 atoms.
(c) Valence charge density and (d) spin density in a plane
perpendicular to the [0001] direction and containing  Mn2 and Ge atoms.
Shading for the
atomic spheres are consistent with Figure 1: white, black and grey circles
denote Ge, Mn1 and Mn2 atoms, respectively. In panels (b) and (d) solid
(dashed) lines show the positive (negative) contribution to the spin
density.}
\vspace{1cm}
\label{chfig}
\end{figure}

\subsection{Hyperfine fields}
We focus next on  hyperfine
fields and compare our predicted values with experimental data\cite{jackson}. 
As is well-known, the hyperfine field\cite{art} of an atom is the magnetic
field at the atomic nuclear site produced by the electrons in the solid and can
be probed using M\"ossbauer spectroscopy or nuclear magnetic resonance to
provide valuable information on the electronic and magnetic properties of the
compound. It
consists of several contributions: (i) the leading
term  due to the Fermi contact interaction 
which is proportional to the
spin density at the nucleus\cite{art},
\begin{equation}
H_{hf}^{ct} = \frac{8}{3} \pi \mu_B^2 [\rho_{\uparrow}(0) - \rho_{\downarrow}(0)],
\label{wf}
\end{equation}
in the scalar
relativistic limit; (ii) an orbital term which, according to Abragam and
Pryce\cite{abrag} can be expressed as:
\begin{equation}
H_{hf}^{orb} \sim 2 \mu_B <r^{-3}>_l \mu^l
\end{equation}
where $<r^{-3}>_l$ is the average expectation value of $r^{-3}$ of the radial
wave function and $\mu^l$ is the orbital magnetic moment;
(iii) a dipolar term $H_{hf}^{dip}$. Whereas only $s$ electrons contribute to
the Fermi contact term, electronic states with $l\neq$0 contribute to the
latter terms. In scalar or non--relativistic calculations, the orbital angular
momentum is quenched and $H_{hf}^{orb}$=0; however, when spin--orbit is
included, this term can be non--vanishing. In our case, the $p$ ($l$ = 1)
contribution to the orbital moment is negligible ($<$10$^{-3}$ $\mu_B$) and the
orbital term will be therefore evaluated only for $l$ = 2 ($d$ states). The
dipolar contribution is normally small in bulk systems and is therefore
neglected.

\begin{table}
\caption{GGA calculated Fermi contact
hyperfine fields  broken down into the core ($H_{hf}^{ct,core}$)
and valence ($H_{hf}^{ct,val}$)
contributions (values in parenthesis denote LSDA calculated values). 
The total Fermi contact hyperfine fields ($H_{hf}^{ct,tot}$), the average
$<r^{-3}>$ for the $d$ states and the orbital hyperfine field ($H_{hf}^{orb}$)
are also shown. The total hyperfine field 
$H_{hf}^{t}=H_{hf}^{ct,tot}+H_{hf}^{orb}$
is compared with
available experimental data (magnitude of 
$H_{hf}^{exp}$)\protect\cite{jackson}. Values of hyperfine fields ($<r^{-3}>$)
are expressed in kOe ($a_0^{-3}$, where $a_0$ is the Bohr radius).}
\begin{tabular}{|c|c|c|c|c|c|c|c|}\hline \hline
 & $H_{hf}^{ct,core}$& $H_{hf}^{ct,val}$ &  $H_{hf}^{ct,tot}$ &
 $<r^{-3}>$ & $H_{hf}^{orb}$ &
 $H_{hf}^t$& $|H_{hf}^{exp}|$ \\ \hline \hline
Mn1 & -316  & 111  & -205  & 2.14 & 12.9& -192 & 195 \\ 
 & (-279) & (76) & (-203) &  & & &  \\ \hline
Mn2 & -459  & 107  & -352  & 2.18 & 9.5 & -342 & 399 \\  
 &  (-417) & (70) &  (-347) & &  &  &  \\ \hline \hline 
\end{tabular}
\label{tabhf}
\end{table}

In Table \ref{tabhf}, we report our calculated values for the 
core and valence contributions to the Fermi contact  hyperfine field, along
with the orbital and total contributions, compared with
  experimental values.
The dominant exchange polarization of
the core electrons has to be taken into account, 
showing the need for an
all--electron method\cite{FLAPW} when dealing with hyperfine fields.
As discussed in previous theoretical work for
transition metals,\cite{art} in \mng the separation  of 
the negative core and  positive valence
contributions to the Fermi contact hyperfine field
highlights these two opposite
terms (cf. Table \ref{tabhf}). 
The large  negative
  core contribution can be attributed to the attraction
of the majority  spin electrons towards the  spatial region
of the spin--polarized $d$
shell\cite{art} which produces the excess of minority spin electrons
   at the  nucleus. 
In order to evaluate the effects of a different parametrization for the
exchange--correlation potential, we compare the Fermi contact term within LSDA
and GGA. The total (core+valence) contribution is very similar; however, both
the separate core and  valence terms have a larger magnitude within GGA. Note
that the core polarization per unit spin moment  for Mn 1 and Mn 2 in both LDA and 
GGA is constant - as is expected
from the exchange polarization mechanism but with a somewhat different constant, 
namely $\sim$ 130 kG/$\mu_B$ and  
$\sim$ 140 kG/$\mu_B$ within LSDA and GGA, respectively. 
These values are pretty similar to the values obtained
for the Mn--based Heusler compounds\cite{heuslv} ($\sim$ 140 kG/$\mu_B$
within LSDA and $\sim$ 150 kG/$\mu_B$ within GGA).

As far as the orbital contribution is concerned, we point out that this term is, as
expected,  much smaller than the Fermi--contact term and very small, 
 because the unquenched orbital moment is so small, as is usual for Mn. Its magnitude could be
slightly underestimated due to the well--known failure of SDFT in determining
orbital magnetic moment; however, this error is expected not  to dramatically
change the final value of the total hyperfine field. In particular, we point
out that the inclusion of the (positive) orbital term improves (worsens) the
agreement with experiment in the case of Mn1 (Mn2).
In fact, 
by means of 
zero--field NMR  and specific
heat measurements, the magnitude of the experimental
effective nuclear fields were determined as 195 kOe at
the 4(d) Mn site and 399 kOe at the 6(g) Mn site\cite{jackson}.
The agreement of the calculated values ($H_{hf}^{tot}$(Mn$_1$) = 
-192 kOe and $H_{hf}^{tot}$(Mn$_2$) = -342 kOe) with experiments is seen to be 
reasonably good.

\section{Fermi velocities and degree of spin--polarization}

Since Mn$_5$Ge$_3$ has been suggested as a potential spin-injector,
it is useful for device applications
to give information about transport properties, in terms of Fermi
velocities and spin-polarization. 

Recall that various definitions of 
spin--polarization $P$ have been proposed, each of them to  be
used in different regimes\cite{prlmazin}. 
The most natural and popular definition involves the DOS at $E_F$ and is probed,
for example, in spin--polarized photoemission measurements: 
\begin{center}
$P_0$ =
[N$_{\uparrow}$($E_F$)-N$_{\downarrow}$($E_F$)]/[N$_{\uparrow}$($E_F$)+N$_{\downarrow}$($E_F$)] 
\end{center}
However, in transport measurements (see for example Andreev
reflection\cite{andreev}),
Fermi velocities are of course relevant quantities and should therefore  be
involved in the spin--polarization definition\cite{prlmazin}. In particular, 
for low resistance ballistic contacts, the appropriate
definition is:
\begin{center}
$P_1$ = ($<N(E_F)\cdot v_F>_{\uparrow}$ - $<N(E_F)\cdot
v_F>_{\downarrow}$)/($<N(E_F)\cdot v_F>_{\uparrow}$ + $<N(E_F)\cdot v_F>_{\downarrow}$)
\end{center}
whereas for large barrier and/or diffusive current the correct definition is:
 \begin{center}
$P_2$ = ($<N(E_F)\cdot v_F^2>_{\uparrow}$ - $<N(E_F)\cdot
v_F^2>_{\downarrow}$)/($<N(E_F)\cdot v_F^2>_{\uparrow}$ + $<N(E_F)\cdot v_F^2>_{\downarrow}$)
\end{center}
where $<>$ denotes the Fermi surface average.

\begin{figure}
\includegraphics[scale=0.4]{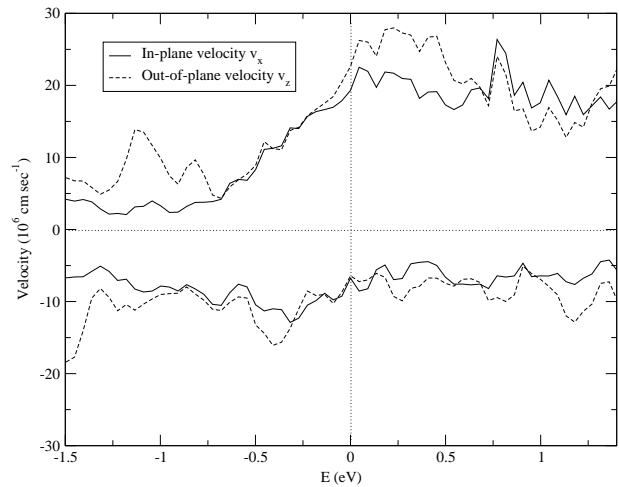}
\caption{In-plane (solid) and out-of-plane 
(dashed) velocities as a function of
energy for majority and minority spin electrons (positive and negative $y$ axis,
respectively).} 
\label{velofig}
\end{figure}

In Figure \ref{velofig} we show our GGA-calculated in-plane (perpendicular to the $c$
axis) and out-of-plane (parallel to the $c$
axis)
velocities as a function of energy.
\begin{table}
\caption{Upper part: GGA-calculated relevant quantities
at E$_F$ for up and down spin-channels
(first and second line, respectively): Density of states
($N(E_F)$, in states/eV), in-plane velocity ($v_{F\parallel}$, 
in 10$^6$ cm/sec), out-of-plane velocity ($v_{F\perp}$, 
in 10$^6$ cm/sec), in-plane plasma frequency ($\omega_{p\parallel}$, in eV)
and out-of-plane plasma frequency ($\omega_{p\perp}$, in eV).
Spin--orbit coupling is not included.
Lower part: Degree of spin--polarization (DSP)
as measured in photoemission measurements ($P_0$),
ballistic transport (in--plane and out--of--plane,
$P_{1\parallel}$ and $P_{1\perp}$, respectively) and  diffusive transport
(in-plane and out-of-plane  ($P_{2\parallel}$ and 
$P_{2\perp}$, respectively.) - see text for definitions. For comparison, we also
show similar quantities obtained for ferromagnetic hcp Co.}  
\begin{tabular}{|c|c|c|c|c|c|c|} \hline \hline
& Compound &$N(E_F)$ & $v_{F\parallel}$ &   $v_{F\perp}$ & $\omega_{p\parallel}$ &
$\omega_{p\perp}$ \\ \hline  
Spin-up & \mng & 3.3 & 19.3 &  22.6 & 2.0 & 2.4 \\
Spin-down &  & 7.9 & 6.7 & 6.4 & 1.1 & 1.1 \\\hline 
Spin-up & Co & 0.3 & 47.9 &  38.2 & 5.0 & 4.0\\
Spin-down &  & 1.5 & 14.0 & 16.4 &  3.3 &  3.8 \\\hline \hline 
&  Compound &$P_0$ & $P_{1\parallel}$ & $P_{1\perp}$ & $P_{2\parallel}$ & $P_{2\perp}$ \\\hline
DSP & \mng & -41$\%$ & 8$\%$  & 18$\%$ & 54$\%$ & 67$\%$ \\\hline 
 & Co & -67$\%$ & -19$\%$  & -37$\%$ & 40$\%$ & 4$\%$ \\\hline 
\end{tabular}
\label{mazin}
\end{table}

Moreover,
in the upper part of Table \ref{mazin}, the
corresponding quantities evaluated at the E$_F$ are reported, along with
 plasma
frequencies.
According to these values, 
we have calculated the  $P_0$, $P_1$ and $P_2$ values 
(the latter two for different in-plane and out-of-plane directions) reported
in Table \ref{mazin} (lower part). Due to the numerical uncertainties (related
to the {\bf k}-point sampling, wave function cut-offs, etc.), we estimate an error
on the spin--polarization of $\pm$5-10 $\%$. Within this error, our values
are consistent with similar values recently obtained using a different DFT method
and slightly different lattice parameters.\cite{pcarerwin}  

As pointed out in the discussion of the electronic properties, the Mn heavy
$d$ bands
in the majority channel are almost fully occupied and the Fermi level 
also crosses Ge
$p$  states, which are light states with an appreciable velocity. On the
other hand, in the minority spin-channels, the DOS shows contributions from both
Mn heavy $d$ and Ge $p$ states.   
As clearly shown in Table \ref{mazin}, the Fermi DOS in
the minority spin--channel is larger - by 
a factor of $\sim$2 - than in the majority spin--channel, 
whereas the Fermi velocities (both in--plane and
out--of--plane) are larger by a factor of $\sim$3 for majority spins. 
This leads to a negative spin-polarization $P_0$, but to a
positive current (see positive values of $P_1$), as also noted in
3$d$ metals.\cite{prlmazin} As for the plasma frequencies, we point
out that there are only slight differences between in-plane and out-of-plane
calculated values, whereas larger differences between
 minority and minority spins emerge: $\omega_p^{\uparrow}$
 is about two times larger than $\omega_p^{\downarrow}$.

  As shown in Figure \ref{velofig},
the anisotropy of the in--plane with
respect to the out--of--plane velocity is quite evident: although the behavior as
a function of  energy is overall similar, they slightly differ in the case of
majority spins, at about -1eV
and in the relevant energy range around the Fermi level (from -0.2 to 0.5 eV
in the majority channel). Moreover, 
as shown in Table \ref{mazin}, the different definitions of spin
polarizations result in largely differing values. 
Both the anisotropy as well as the differences
among $P_0$, $P_1$ and $P_2$ should help in exploiting this compound
for spin--injections purposes 
in the most appropriate 
transport 
regime and along the most favorable growth direction. 
In particular, we remark that with
$P_{2\perp}\sim$ 70\%, most of the current along the $c$ axis in the
diffusive (Ohmic) regime is therefore
carried by
 majority spins. 
 Of course, this picture might be modified in the presence of
 a junction (such as Mn$_5$Ge$_3$/Ge), where interface states can modify
 the electronic structure and velocities with respect to the ideal bulk
 situation considered here.

Finally, it is useful to compare calculated
spin-injection efficiencies and Fermi velocities
for \mng with those obtained for ferromagnetic hcp Co\cite{notaco}, 
a widely used material in high 
density magnetic recording media. Our results are shown in
 Table \ref{mazin}. Interestingly, the cobalt Fermi velocities (and related
 plasma frequencies) are larger by a factor of 2-3 compared to \mng, also
 showing a larger anisotropy. As far as the Co DSP is concerned, we remark that,
 similarly to \mng~,
 $P_0$ is negative; however,  $P_1$ - both in--plane and out--of--plane
 - is also negative, at variance with Mn$_5$Ge$_3$. 
 As a final remark, we observe that
 the Co spin polarization in the ballistic regime is slightly
 higher than in \mng, whereas  it is definitely lower in the diffusive regime, 
 holding
 promise for \mng as an efficient spin--injector.

\section{X--ray Absorption and Magnetic Circular Dichroism}

\begin{figure}
\includegraphics[scale=0.45]{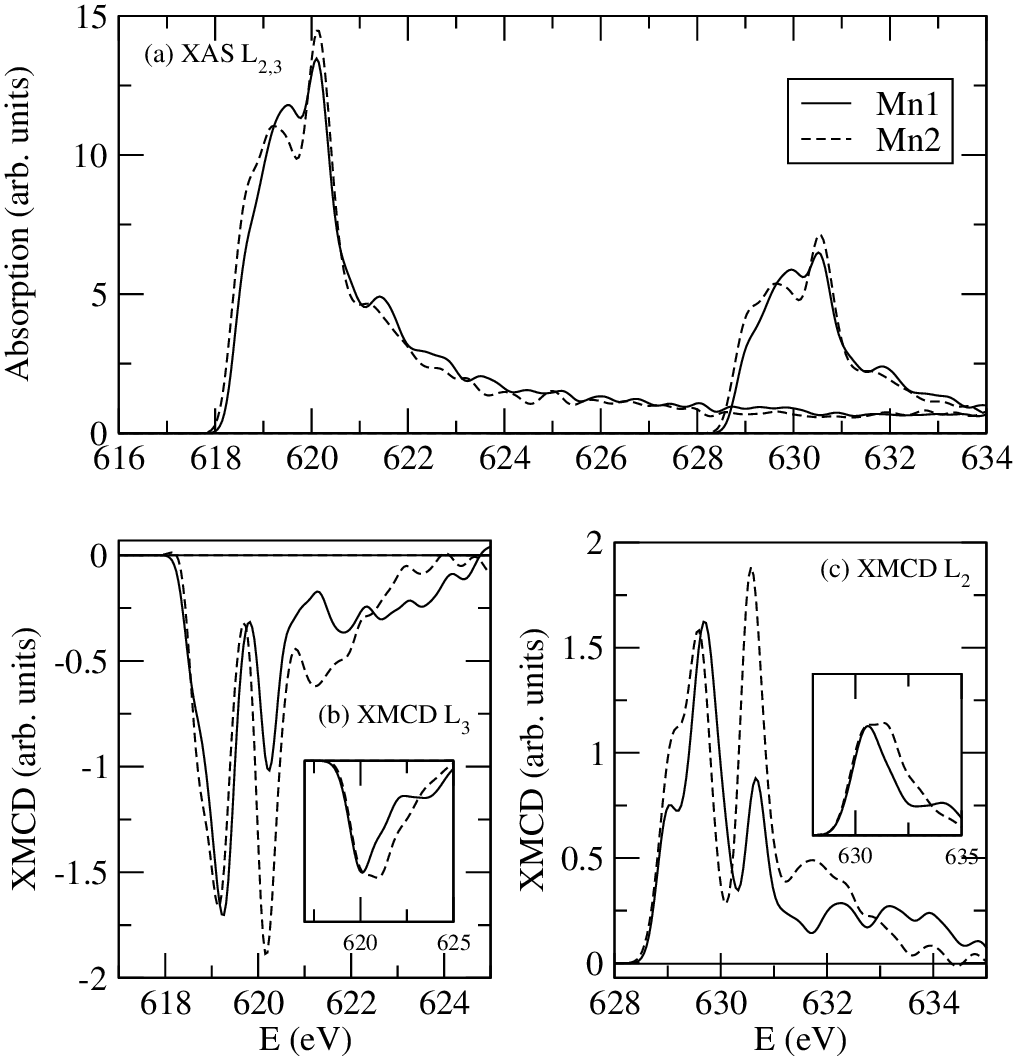}
\caption{(a) $L_{2,3}$ X-ray absorption; 
 (b)  $L_3$ X-ray MCD; and (c) $L_2$  X-ray MCD. The insets in panels
 (b) and (c) show MCD spectra convoluted with an energy broadening
 of 0.7 eV. In all panels,
the solid (dashed) line shows the contribution of the Mn1 (Mn2) site.}
\label{xmcd}
\end{figure}

The formalism within band theory to calculate the cross section for the absorption of incident light
is discussed in Ref.\onlinecite{wu} 
 and is briefly reviewed here. 
For dipole--excited transitions, the cross section
can be obtained as:

\begin{equation}
\sigma_n(E) = \int_{\Omega_{BZ}} |<\Psi_c|p|\Psi_v>|^2 \delta(E-(E_v-E_c))
d{\bf k}
\end{equation}
where $n$ = $z$, $\pm$ represents the photon polarization  (i.e.
incident light polarized vertically along the direction of magnetization
($z$) or left- ($+$) or right--circularly ($-$) polarized),
$p$ is the momentum operator, $\Psi_c$ and $\Psi_v$ ($E_c$ and $E_v$)
denote initial core
and final valence states (eigenenergies), respectively.
Therefore, the X-ray 
magnetic circular dichroism (XMCD) can be obtained as
$\sigma_m = \sigma_+ - \sigma_-$, whereas the X-ray absorption spectrum
(XAS) is calculated as $\sigma_t = \sigma_+ +
 \sigma_- + \sigma_0$ ($\sigma_0$ denotes the absorption cross section for
 incident light polarized vertically along the direction of magnetization, here
 chosen as coincident with the $z$ hexagonal axis).
Starting from the converged FLAPW ground
state, the spin-orbit coupling (SOC) was treated in a second
variational way\cite{macsoc} to obtain
the XAS and XMCD spectra, using up to 432 {\bf k}-points in the full
Brillouin zone. We used a 0.25 eV Lorentzian broadening to 
smooth the calculated
spectra, in order to take lifetime effects into account (see below). 

The energy dependence of $L_{2,3}$ $\sigma_m$ and   $\sigma_t$ are shown in
Figure \ref{xmcd} for the two Mn atomic types.
As is usual for 3$d$ metals,\cite{wu} and pointed out above in the PDOS
discussion,  most of the majority spin bands are
located below $E_F$, so that photon induced transitions occur mainly to the unoccupied
minority spin--bands.
We haven't included any self--energy correction, so we expect the calculated binding
energy of the 2$p$ states to be underestimated by several tenths of an eV. 
The energy difference  between the $L_2$ and $L_3$ edge represents the size of the
spin-orbit splitting of the 2$p$ core states, and is usually found to be in good agreement with
experiment. For \mng it is estimated to be $\Delta_{SOC}^{2p}$
= 10.4 eV. This value is similar to other theoretical results for 
Mn based alloys, such as the  Heusler
PtMnSb and NiMnSb\cite{galan}, as well as to other experimental data obtained
for DMS\cite{jonker}.
As expected from the quite low symmetry of the hexagonal lattice 
(and the related difference of the out--of--plane
$z$ direction compared to the in--plane $x,y$ direction), $\sigma_0$ is quite
different from  $\frac{1}{2}(\sigma_+ + \sigma_-)$ (not given in
 Figure \ref{xmcd}).
The XAS spectrum shows tails extending to high energy ($>$10-15 eV)
with respect
to the main absorption peaks, whereas the XMCD becomes almost
negligible at 5-7 eV
above the absorption edge. For the energy broadening
 value used (0.25 eV), both the absorption and
dichroism spectra show a quite  rich structure; in particular, 
the XAS and - even more markedly - the XMCD spectra
show a double peak structure (related
to the peculiar features
in the unoccupied density of states, see Fig.\ref{pdos} (b) and (c)), 
followed by a smaller bump at about 3.5 eV above
the absorption edge. 
As far as the comparison between Mn1 and Mn2 is concerned, we remark that the
XAS spectrum shows similar features, whereas the XMCD shows different
amplitudes of the second peak, which is more marked in Mn2 compared to Mn1.
 This difference in the XMCD
 intensity is consistent with the larger Mn2 magnetic moment compared to Mn1.
Since the broadening used
(0.25 eV) is lower
than the common
experimental resolution, 
we also show in the insets of Fig.\ref{xmcd} (b) and (c) the same XMCD
spectra obtained using a Lorentzian broadening equal to 0.7 eV; in this case, the
double--peak feature cannot be resolved, whereas
 the larger amplitudes observed for Mn2
compared to Mn1 in the high energy range is still evident.

 To further investigate this issue, we recall that
 some important magneto-optical sum rules have been derived in
 recent years, which relate
 the integrated signals over the spin--orbit split
 core edges of the unpolarized XAS and of  circular dichroism to ground--state
 orbital and spin magnetic moments\cite{vanderlaan,carra,wuf,solids}. 
The orbital and spin sum rules are expressed as:
\begin{center}
\begin{eqnarray}
<l_z> = \frac{2 I_m N_h}{I_t} \\
<s_z> = \frac{3 I_s N_h}{I_t} - 7 <T_z> \\
I_m = \int [(\sigma_m)_{L_3} + (\sigma_m)_{L_2}] d\epsilon \\
I_s = \int [(\sigma_m)_{L_3} - 2(\sigma_m)_{L_2}] d\epsilon \\
I_t = \int [(\sigma_t)_{L_3} + (\sigma_t)_{L_2}] d\epsilon 
\end{eqnarray} 
\end{center}
where $N_h$ is the number of
 holes in the $d$ band, $N_h = 10 - n_{3d}$ (with $n_{3d}$ determined by
 the $d$ projected density of states inside each atomic sphere).
$T_z$
is the $z$ component of the magnetic dipole operator:
\begin{center}
$T_z$ = 1/2 [{\boldmath $\sigma$} 
- 3 $\hat{\bf r} (\hat{\bf r}\cdot$ {\boldmath $\sigma$})$]_z$
\end{center} 
with {\boldmath $\sigma$}  denoting the vector of Pauli matrices and
is related to the non-spherical charge and spin density.
 $T_z$ is  vanishing for cubic systems, whereas it is not necessarily
negligible in hexagonal systems. However, 
preliminary calculations suggest that $T_z$ is pretty small 
(at most of the order of 0.03)
 and is therefore neglected
 here.  The orbital and spin magnetic moments are 
 $\mu^l = -\mu_B <l_z>$ and $\mu^s = -\mu_B <s_z>$, respectively. 
Their ratio, as derived from sum rules, reads as:
\begin{equation}
\frac{\mu^l}{\mu^s}=\frac{<l_z>}{<s_z>}= \left[ \frac{3\: I_s}{2\: I_m} - \frac{7<T_z> I_t}{2 \:I_m \:N_h} \right]^{-1}
\end{equation} 
 
Experimentally, since $<T_z>$ is very hard to access,  it is generally
neglected when considering sum rules; in this case the experimental
uncertainties - related to fixing somehow the number of holes $N_h$ or to 
calculating $I_t$ - drop out. It has therefore been suggested \cite{carra2}
that an accurate estimate of the $\mu^l/\mu^s$ ratio can be obtained from the
$2 I_m/3 I_s$ ratio.
 
\begin{table}
\caption{Orbital and spin magnetic moments as determined from sum--rules (SR)
and
self-consistently (SC) for Mn1 and Mn2
atoms along with their ratio.}
\begin{tabular}{|c|c|c|c|c|c|c|}  \hline \hline
& \multicolumn{2}{c|}{$\mu^l$} & \multicolumn{2}{c|}{$\mu^s$}&
\multicolumn{2}{c|}{$\mu^l/\mu^s$} \\  \hline
& SR & SC & SR & SC & SR & SC  \\ \hline \hline
Mn1 & 0.03 & 0.05 & 1.9 & 2.07 & 0.015 & 0.024 \\ \hline
Mn2 & 0.02 & 0.035 & 2.4 & 3.16 & 0.01 & 0.01 \\ \hline \hline
\end{tabular}
\label{sumrule}
\end{table}

 In Table \ref{sumrule} we report our calculated values for the orbital and spin
 magnetic moments, as well as their ratio,
 as determined from sum rules (SR) for Mn1 and Mn2 atoms. For comparison, we also
 report the same values as self-consistently calculated (cfr. Table
\ref{tabmom}).   
 As expected, the general trends of spin and magnetic moments for Mn1 and
 Mn2 atoms
 are  qualitatively well reproduced. However, quantitatively, the SR 
magnetic moments
 are generally
 underestimated with respect to the self-consistently calculated values
 (ranging from $\sim$10 percent in the
 case of spin magnetic moments up to 40 $\%$ in the case of 
the small orbital moments).
 This has been  ascribed\cite{wuf}  to the several approximations made in deriving
 the sum rules\cite{ebert}, among which the most serious are {\em i}) ignoring the
 interatomic hybridization,
 {\em ii}) neglecting
 the $p\rightarrow s$ transitions\cite{wuf}, and { \em iii})  ignoring the exchange splitting
 of core levels. Therefore, first--principles calculations of both XMCD spectra
 and
 ground--state magnetic moments are crucial for a careful study of the \mng
 compound  and, eventually, for a quantitative interpretation
 of future experimental results. 
 
\section{Summary}
\label{concl}

In the search for new compounds to be used for efficient
spin--injection in spintronic
devices and
following the recent suggestion of Mn$_5$Ge$_3$/Ge(111) system as a promising 
system, we have presented a careful first-principles FLAPW
investigation of the electronic, transport,
magnetic and magneto--optical properties of bulk Mn$_5$Ge$_3$.
Our results show that the two Mn sites have different magnetic moments,
leading to a total
magnetization of 26 $\mu_B$ per  unit cell; the conducting character is strongly
metallic with states around the Fermi level essentially due to Mn-Mn interactions.
 Our theoretical predictions were carefully analyzed in
terms of the underlying electronic and magnetic structure and shown to be  in
 excellent quantitative and
qualitative agreement with available experimental results.
The most favorable condition for spin--injection purposes is predicted to be
in the diffusive regime along the hexagonal $c$ axis, where a rather high
spin-polarization  is obtained.

\acknowledgments{We thank Prof. Ruqian Wu and Prof.
Sandro Massidda for assistance
provided with the calculation of the magnetic dipole operator
and spline fitting procedure, 
respectively. Useful discussions with Dr. Steven C. Erwin are
gratefully acknowledged.
Work in L'Aquila
supported by INFM through Iniziativa Trasversale Calcolo
Parallelo and PAIS-GEMASE project.}  

\newpage

\begin{figure}
\includegraphics[scale=0.2]{chfig.epsi}
\end{figure}


\begin{thebibliography}{100}
\bibitem{notaco} The hcp Co was simulated using the experimental lattice constants
$a$ = 2.51 \AA and $c$ = 4.07 \AA.
\end{thebibliography}
\end{document}